\documentclass[12pt,preprint]{aastex}

\usepackage{CJK}
\begin{document}
\begin{CJK}{UTF8}{gbsn}

\title {$^{26}$Al in the Early Solar System: Not so Unusual After All}

\author{M. Jura\altaffilmark{a}, S. Xu\altaffilmark{a}(许\CJKfamily{bsmi}偲\CJKfamily{gbsn}艺), \& E. D. Young\altaffilmark{b}}

\altaffiltext{a}{Department of Physics and Astronomy, University of California, Los Angeles CA 90095-1562; jura, sxu@astro.ucla.edu}
\altaffiltext{b}{Department of Earth and Space Sciences, University of California, Los Angeles CA 90095; eyoung@ess.ucla.edu}

\begin{abstract}
Recently acquired evidence shows that extrasolar asteroids exhibit over a factor of 100 variation in the  iron to aluminum abundance ratio.  This large range likely is a  consequence of  igneous differentiation that resulted from heating produced by radioactive decay of $^{26}$Al with an abundance comparable to  that  in the solar system's protoplanetary disk at birth. If so, the conventional view that our solar system began with an unusually high amount of $^{26}$Al should be discarded.
  \end{abstract}
\section{INTRODUCTION}

 Concentrations of excess $^{26}$Mg, the decay product of the short-lived radionuclide $^{26}$Al [mean life = 1.03 Myr \citep{Castillo-Rogez2009}], show that the solar system 
 formed with $n$($^{26}$Al)/$n$($^{27}$Al) = 5.2 ${\times}$ 10$^{-5}$ \citep{Jacobsen2008}.  
 Although there is evidence that there may have been deviations from this ``canonical" ratio across the solar protoplanetary disk by as much as a factor of 2 \citep{Larsen2011,Liu2012},  the overall concentration of $^{26}$Al in the solar disk was more than  a factor of 10 greater than the current average value in the interstellar medium of 3.0 ${\times}$ 10$^{-6}$ \citep{Tang2012}.
   While some $^{26}$Al may have been produced within the early solar system, most of it was not \citep{Duprat2007, Desch2010}; there must have been a significant external source of this short-lived nuclide.   Commonly, the natal $^{26}$Al is taken as a signature of a nearby supernova that may have triggered the collapse of the molecular cloud from which the Sun formed \citep{Meyer2000, Gritschneder2012}.  Alternatively, winds from massive stars may have supplied the bulk  of the 
 $^{26}$Al \citep{Prantzos2004, Gaidos2009, Gounelle2012}. 
 
A major consequence of  large amounts of $^{26}$Al in the early solar system was substantial internal heating of young planetesimals which therefore melted and subsequently experienced igneous differentiation.    Iron meteorites
 are thought to be modern fragments  of iron-rich cores formed during this era \citep{McSween2010}.  If other planetary systems  formed with considerably less $^{26}$Al, then their asteroids may
 not be differentiated.  We can test this scenario by examining the elemental compositions of extrasolar minor planets. 

Evidence is now compelling that some white dwarfs have accreted some of their own asteroids \citep{Debes2002, Jura2003, Jura2014}.  In some instances, we have detected excess infrared emission from circumstellar disks
composed of dust  \citep{Farihi2009, Xu2012} where gas also is sometimes evident \citep{Gaensicke2006}.  These disks lie within the tidal radius of the white dwarf and are
understood to be the consequence of an asteroid having been shredded after its orbit was perturbed so it passed very close to the star \citep{Debes2002, Bonsor2011, Debes2012}.  Accretion from these disks supplies the orbited white dwarf's atmosphere with elements heavier than helium  where they are normally not found because the
gravitationally settling  times are very short compared to the cooling age of the star. Estimates  of the amount of accreted mass argue that we are witnessing
the long-lived evolution of ancient asteroid belts \citep{Zuckerman2010, Jura2014}. In the most extreme case, the accreted parent body may have been as massive as Ceres \citep{Dufour2012} which has a radius near 500 km.  However, the required mass more typically implies parent bodies with radii near 200 km \citep{Xu2013}.   Externally-polluted white dwarfs  provide a means for placing the solar concentration of $^{26}$Al in context. 
 
 As a first approximation, extrasolar asteroids resemble bulk Earth being largely composed of oxygen, magnesium, silicon and iron and deficient
 in volatiles such as carbon and water \citep{Klein2010, Jura2006, Jura2012a} as expected in simple models for planet formation from a nebular disk.  When  eight or more polluting elements are detected, it is possible
 to tightly constrain the history and evolution of the parent body \citep{Zuckerman2007, Xu2013}.  Recent studies of such richly polluted stars have shown abundance patterns that can  be best explained if  the accreted planetesimal evolved
 beyond simple condensation from the nebula where it formed.  For example, NLTT 43806 is aluminum rich as would be expected if the accreted planetesimal
 largely was composed of a crust \citep{Zuckerman2011} while PG 0843+516 is iron rich which can be explained by the accretion of a core \citep{Gaensicke2012}.  \citet{Xu2013} found that the abundance pattern of the object accreted onto GD 362 resembles that of a mesosiderite -- a rare
 kind of meteorite that is best understood as a blend of core and crustal material \citep{Scott2001}.  
  
 Here, we first revisit the current sample of extrasolar planetesimals with well-measured abundances and reconfirm that igneous differentiation is
 widespread \citep{Jura2014}.  We then present a model to explain this result.  Finally, we consider our solar system from the perspective of  extrasolar
 environments.

 \section{EVIDENCE FOR WIDESPREAD DIFFERENTIATION}

The evidence for igneous differentiation among extrasolar planetesimals can be presented in a variety of ways \citep{Jura2014}.  Here,  
we display in Figure 1 the abundance ratios by number, $n$(Fe)/$n$(Al) vs. $n$(Si)/$n$(Al),  for all seven  externally-polluted white dwarf atmospheres where these three elements have been reported\footnote{The stars are GD 362 \citep{Xu2013}, GD 40 \citep{Jura2012b}, NLTT 43806 \citep{Zuckerman2011},
PG 0843+516, WD 1226+110 and WD 1919+012 \citep{Gaensicke2012} and WD J0738+1835 \citep{Dufour2012}.  All but NLTT 43806 harbor a dust disk while circumstellar gas has been detected orbiting PG 0834+516 and WD 1226+110.}.  We see that $n$(Fe)/$n$(Al) varies
by more than a factor of 100, a much greater range than shown among main-sequence planet-hosting stars, solar system chondrites and even $n$(Si)/$n$(Al) among these same polluted stars.  
\begin{figure}
 \plotone{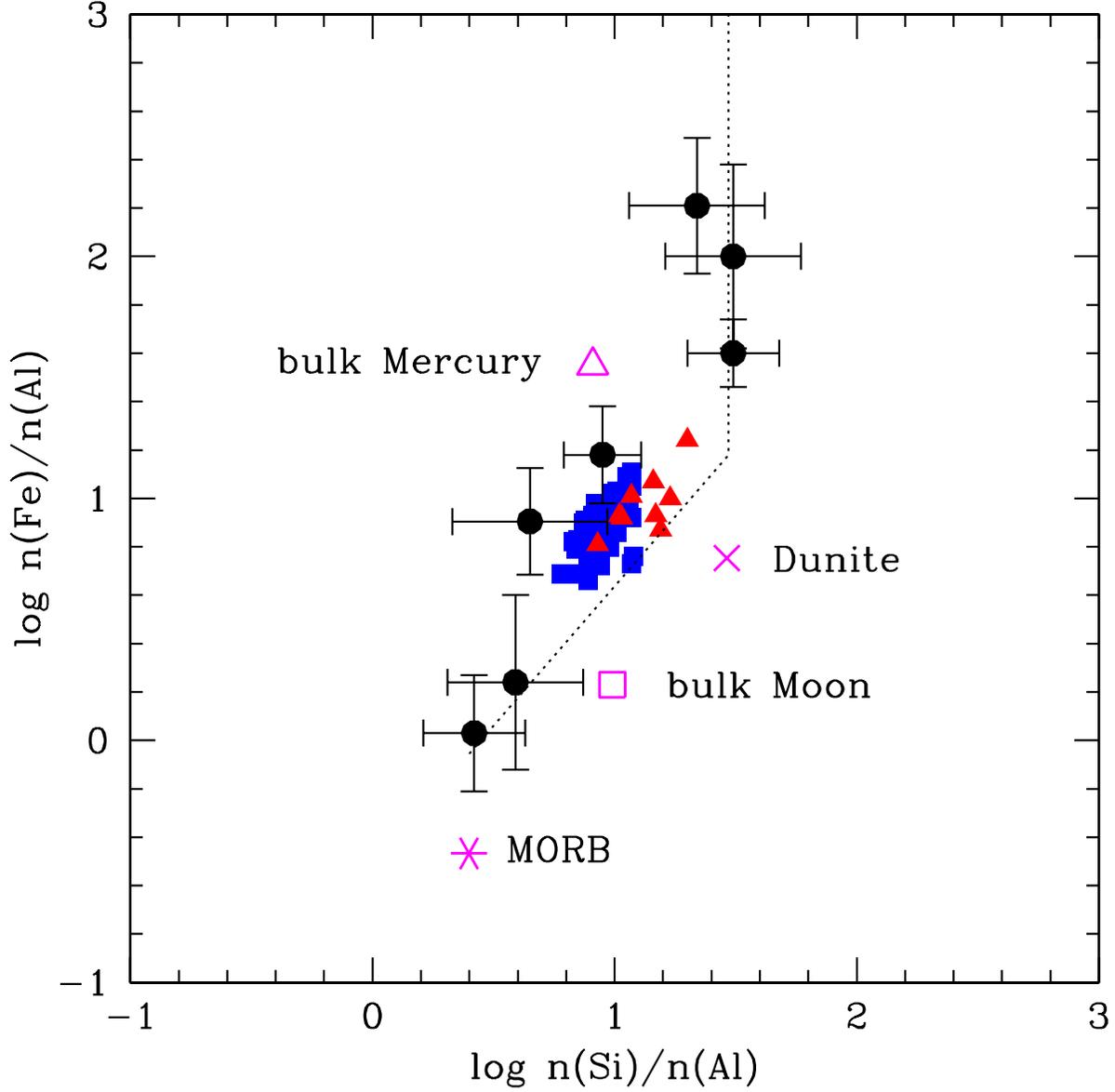}
\caption{Abundance ratios by number of $n$(Fe)/$n$(Al) vs. $n$(Si)/$n$(Al) are denoted by black circles for those seven systems 
 where all three elements have been detected.  We assume a steady state approximation where the element's mass in the star's mixing zone is governed by a balance between the rate at which atoms are accreted and  
  the rate at which they settle gravitationally into the interior \citep{Jura2014}.   For comparison, we also display abundances among solar system chondrites as red triangles \citep{Wasson1988} and planet-hosting stars as  blue squares \citep{Gilli2006}.  Solar system materials are displayed by magenta symbols. The vertical dotted line represents model planetesimals composed  of a blend of core and mantle rocks.  The sloped dotted line represents model planetesimals composed of a blend of mantle and crustal rocks  plus a  core with 10\% of the total mass.  The observed ratios in externally-polluted white dwarfs can be
  reproduced with different combinations of solar system objects.} 
\end{figure}
 
The large range in $n$(Fe)/$n$(Al) among extrasolar planetesimals must be the result of  some powerful cosmochemical process. 
 One  possibility is that unlike in the solar system,
some extrasolar planetesimals were formed largely of refractory elements \citep{Bond2010} resulting in  low values of $n$(Fe)/$n$(Al) because Al is highly refractory .  However, this scenario
is  not supported by available observations \citep{Jura2013}, and cannot explain why some systems have relatively high values of $n$(Fe)/$n$(Al).  Because there is
no viable nebular model to explain the observed range in $n$(Fe)/$n$(Al), the abundance variations must have been produced
within the planetesimals themselves.  

Abundance patterns in extrasolar planetesimals reproduce those in familiar rocks.  The lowest value of $n$(Fe)/$n$(Al) is comparable to the ratio in MORB (Mid Ocean Ridge 
Basalt), a characteristic crustal rock \citep{Presnall1987}.  The highest value of $n$(Fe)/$n$(Al) exceeds that of dunite, a  mantle rock, implying sampling of iron-rich core material \citep{Hanghoj2010}.   Figure 1 shows that the range of $n$(Fe)/$n$(Al) among extrasolar asteroids is even greater than the difference found between  bulk Moon \citep{Warren2005} and bulk Mercury \citep{Brown2009}, two solar system objects which are understood as having a small
and large iron core, respectively.  

We understand the variety of elemental compositions among extrasolar planetesimals as the  consequence of a familiar three-step process.  First, planetesimals form within the disk; in this environment, volatiles such as water may be excluded.  Second, differentiation 
results in iron being concentrated
in the core  and aluminum being  concentrated in the crust.  Third, collisions  lead to stripping and blending of cores and crusts with a consequent
dramatic variation in $n$(Fe)/$n$(Al) in the end-product planetesimals, blends of  different portions  of core, mantle and crustal material.  

\section{MODEL}
As with solar system asteroids,  the heat source for igneous differentiation of extrasolar planetesimals most likely was from radioactive decay of $^{26}$Al \citep{Ghosh1998}.   Other possibilities do not seem viable.  The gravitational potential energy   released  by forming a body of radius, $R_{0}$, and mass, $M$, can raise the temperature an amount, ${\Delta}$T given by:
\begin{equation}
C\,M\,{\Delta}T\;=\;\frac{3}{5}\,\frac{G\,M^{2}}{R_{0}},
\end{equation}
where $C$ is the specific heat (J kg$^{-1}$ K$^{-1}$) and $G$ is the gravitational constant.  For a typical object with $R_{0}$ ${\approx}$ 200 km and $M$ ${\approx}$ 1.0 ${\times}$ 10$^{20}$ kg \citep{Jura2014} and with $C$ ${\approx}$ 1000 J kg$^{-1}$ s$^{-1}$ \citep{Turcotte2002}, then ${\Delta}T$ ${\approx}$ 20 K, much too small to be of importance.
Although mutual collisions 
can produce local heating, it seems unlikely that most of the material is melted during the period of planetesimal
growth by collisions \citep{Davison2010}.   
Within the  average interstellar medium,  $^{60}$Fe/$^{56}$Fe =  2.8 ${\times}$ 10$^{-7}$ \citep{Tang2012}, and if this ratio
prevails within star forming regions, then heating from the radioactive decay of $^{60}$Fe cannot be an important heating source within extrasolar planetesimals.  It is possible that some stars form near supernovae that produce large amounts of $^{60}$Fe \citep{Vasileiadis2013}, and in these environments newly formed planetesimals could be  significantly heated by radioactive decay of this
radionuclide.  However, because by number there is more $^{26}$Al than  $^{60}$Fe within the entire Galaxy \citep{Tang2012} and because $^{26}$Al is
readily produced within massive stars and then injected into the local molecular interstellar medium where new stars form \citep{Gounelle2012}, it is probable that the majority of young stellar
disks are similar to our own solar system where radioactive decay of $^{26}$Al was the dominant source of planetesimal heating.

The usual expression \citep{Turcotte2002} governing the time ($t$) variation of internal temperature, $T$, as a function of radius, $r$, of a spherical rocky body is:
\begin{equation}
\frac{{\partial}T}{{\partial}t}\;\;=\;\frac{1}{r^{2}}\frac{{\partial}}{{\partial}r}\,\left(r^{2}\,{\kappa}\,\frac{{\partial}T}{{\partial}r}\right)\;+\;\frac{{\dot Q}(t)}{C(T)}
\end{equation}
where ${\kappa}$ (m$^{2}$ s$^{-1}$) is the thermal diffusivity and ${\dot Q}(t)$ (J kg$^{-1}$ s$^{-1}$) is the heating energy per unit mass per unit time. The typical timescale for the loss of internal heat is $R_{0}^{2}\,{\kappa}^{-1}$. For a 200 km radius object with a thermal diffusivity of 10$^{-6}$ m$^{2}$ s$^{-1}$ \citep{Turcotte2002},  the  outward diffusion of heat represented by the first term on the right hand side of Equation (2) typically requires more than 1 Gyr  and has a negligible effect on the body's central temperature during the era of heating from $^{26}$Al.  To compute the maximum internal temperature, $T_{Max}$, we integrate over a time scale much longer than the average decay time, $t_{R}$,  and consider the total released energy, $Q_{0}$, defined as:
\begin{equation}
Q_{0}\;=\;{\int}_{0}^{\infty}{\dot Q}(t)\,dt
\end{equation}
from the decay of $^{26}$Al.  Consequently, if the planetesimal originates at temperature, $T_{0}$ (K),  then:
\begin{equation}
{\int}^{T_{max}}_{T_{0}}C(T)\,dT\;{\approx}\;f_{26}\,Q_{0}
\end{equation}
where $f_{26}$ is the initial fraction of the mass of the planetesimal that is $^{26}$Al.  

We assume that igneous differentiation is only possible if the internal temperature exceeds the solidus temperature, $T_{solidus}$, \citep{Turcotte2002} and then derive the
minimum aluminum isotope ratio by number, $n(^{26}$Al)/$n(^{27}$Al), required to achieve this temperature. 
We assume that the extrasolar planetesimal will have formed at some time, $t_{form}$, after inheritance of $^{26}$Al from the molecular cloud.   Subsequently,   no fresh
$^{26}$Al  enters the star-forming cloud;  instead, there is only radioactive decay with mean life, $t_{R}$.  Validated by our detailed calculations not shown here, we take $C$ to be constant and independent of temperature.  If $f_{Al}$ is the fraction of mass of the planetesimal
which is aluminum, and if $T_{0}$ $<<$ $T_{solidus}$, then:
\begin{equation}
\left(\frac{n(^{26}Al)}{n(^{27}Al)}\right)\;{\geq}\;\;\left(\frac{27}{26}\right)\,\left(\frac{T_{solidus}\,C}{Q_{0}\,f_{Al}}\right)\,e^{t_{form}/t_{R}}\;.
\end{equation}

Using CV chondrites with their relatively high Al, thus providing a minimum for Equation [5], we take $f_{Al}$ = 0.0175 \citep{Wasson1988}.  We adopt $T_{solidus}$ =1500 K  \citep{Turcotte2002}, and,  for $^{26}$Al, we take $Q_{0}$ =  1.2 ${\times}$ 10$^{13}$ J kg$^{-1}$ \citep{Castillo-Rogez2009}.  We assume two contributions to $t_{form}$.     First, there is  free-fall gravitational collapse  of a cloud core with an initial radius of 0.1 parsec  that requires ${\approx}$ 0.5 Myr \citep{Hartmann2009}.  Second, planetesimals must assemble within the disk which, by analogy with the solar system, probably takes ${\sim}$1 Myr \citep{Zhou2013}.  Adding both terms, $t_{form}$ = 1.5 Myr.  Consequently, we compute from Equation (5)
that  in extrasolar environments where planetesimals internally melted, $n(^{26}$Al)/{$n(^{27}$Al)} ${\geq}$  3 ${\times}$ 10$^{-5}$, approximately its value in the early solar system.  This result is inexact.  If, for example, we take $f_{Al}$ = 0.0086 as found in CI chondrites \citep{Wasson1988}, then the minimum values of $n(^{26}$Al)/{$n(^{27}$Al)} should
be  doubled.

\section{DISCUSSION AND PERSPECTIVE}
As has been previously suggested qualitatively,  a general enrichment of  $^{26}$Al in protoplanetary disks might occur if  this radionuclide
 is not distributed evenly throughout the Milky Way but, instead, is
confined  to regions of star formation \citep{Draine2011}. A plausible model to explain why $^{26}$Al would be so concentrated  is that this species is largely injected into the interstellar medium from rotating massive stars \citep{Gaidos2009,Gounelle2012}.  
These massive stars are so short lived, that they all reside near their birth sites within molecular clouds.  Such a model can also explain why the solar system has a relatively high concentration of $^{26}$Al
and a relatively low concentration of $^{60}$Fe \citep{Tang2012}. 
However,  winds from Wolf-Rayet stars might shred cloud cores and prevent the formation of planets \citep{Boss2013}.  While some recent models for supernova ejecta into  molecular clouds also predict that solar mass  stars commonly 
form with elevated amounts of $^{26}$Al \citep{Pan2012, Vasileiadis2013}, they do not naturally explain the solar system's simultaneously  depressed value of $^{60}$Fe/$^{56}$Fe.  

 In the entire Milky Way, the mass of
hydrogen  in H$_{2}$  is 8.4 ${\times}$ 10$^{8}$ M$_{\odot}$ \citep{Draine2011}.  The amount of interstellar $^{26}$Al is measured from the intensity of the ${\gamma}$-ray line at 1.8 MeV that results from its radioactive decay.  Including foreground emission, there is somewhere between 1.5 and 2.2 M$_{\odot}$  of  $^{26}$Al within the Galaxy \citep{Martin2009}. If we assume the solar aluminum abundance of
$n$($^{27}$Al)/$n$(H) = 3.5 ${\times}$ 10$^{-6}$ \citep{Lodders2003}, and if  all  measured interstellar $^{26}$Al is confined only to molecular clouds, then in these locations
$n(^{26}$Al)/$n(^{27}$Al) ${\approx}$ 2.0 - 3.0 ${\times}$ 10$^{-5}$, nearly the same as the minimum ratio we infer for the birth environment of extrasolar planetesimals.  Consider not only the entire Milky Way but also  observations of the Orion region,
the nearest molecular cloud where large numbers of high-mass stars currently are being formed.  Orion's ${\gamma}$-ray line emission is explained with 5.8 ${\times}$ 10$^{-4}$ M$_{\odot}$ of $^{26}$Al \citep{Voss2010} from a region where the total mass of H$_{2}$ is approximately 2 ${\times}$ 10$^{5}$ M$_{\odot}$ \citep{Genzel1989}.  The implied value
of $n(^{26}$Al)/$n(^{27}$Al) in the Orion star-forming region is therefore 3 ${\times}$ 10$^{-5}$, again substantially elevated over the average interstellar value.
Remarkably, the apparent fraction of $^{26}$Al within star-forming molecular clouds agrees with the value required to explain the widespread occurrence of differentiated 
extrasolar planetesimals.   It follows that the solar system's initial complement of $^{26}$Al was essentially normal.

We thank L.  Yeung for suggesting a title for this manuscript.  This work has been partly supported by the NSF.  

 \bibliographystyle{apj}

\begin{thebibliography}{99}
\bibitem [{{Bond} {et~al.}(2010)}] {Bond2010} Bond, J. C., O'Brien, D. P., \& Lauretta, D. S. 2010, \apj, 715, 1050
\bibitem [{{Bonsor} {et~al.}(2011)}] {Bonsor2011} Bonsor, A., Mustill, A. J., \& Wyatt, M. C. 2011, \mnras, 414, 930
\bibitem [{{Boss} \& {Keiser}(2013)}] {Boss2013} Boss, A. P. \& Keiser, S. A. 2013, \apj, 770, 51
\bibitem [{{Brown} \& {Elkins-Tanton}(2009)}] {Brown2009} Brown, S. M., \& Elkins-Tanton, L. T. 2009, E\&PSL, 286, 446
\bibitem [{{Castillo-Rogez} {et~al.}(2009)}] {Castillo-Rogez2009} Castillo-Rogez, J., Lee, M. H., Turner, N. J., et al. 2009, Icar., 204, 658
\bibitem [{{Davison} {et~al.}(2010)}] {Davison2010} Davison,T. M., Collins, G. S., \& Ciesla, F. J. 2010, Icar., 208, 468
\bibitem [{{Debes} \& {Sigurdsson}(2002)}] {Debes2002} Debes, J. H., \& Sigurdsson, S. 2002, \apj, 572, 556
\bibitem [{{Debes} {et~al.}(2012)}] {Debes2012} Debes, J. H., Walsh, K. J., \& Stark, C. 2012, \apj, 747, 148
\bibitem [{{Desch} {et~al.}(2010)}] {Desch2010} Desch, S. J., Morris, M. A., Connolly, H. C., \& Boss, A. P. 2010, \apj, 725, 692
\bibitem [{{Draine} (2011)}] {Draine2011} Draine, B. T. 2011, Physics of the Interstellar and Intergalactic Medium (Princeton: Princeton Univ. Press)
\bibitem [{{Dufour} {et~al.}(2012)}] {Dufour2012} Dufour, P., Kilic, M., Fontaine, G. et al., 2012, \apj, 749, 6
\bibitem [{{Duprat} \& {Tatischeff}(2007)}] {Duprat2007} Duprat, J. \& Tatischeff, V. 2007, \apj, 671, L69
\bibitem [{{Farihi} {et~al.}(2009)}] {Farihi2009} Farihi, J., Jura, M., \& Zuckerman, B. 2009, \apj, 694, 805
\bibitem [{{Gaensicke} {et~al.}(2012)}] {Gaensicke2012} Gaensicke, B., T., Koester, D., Farihi, J., et al.,  2012, \mnras, 424, 323
\bibitem [{{Gaensicke} {et~al.} (2006)}] {Gaensicke2006} Gaenscike, B T., Marsh, T. R., Southworth, J., \& Rebassa-Mansergas, A. 2006, Science, 314, 1908
\bibitem [{{Gaidos} {et~al.}(2009)}] {Gaidos2009} Gaidos, E., Krot, A. N., Williams, J. P., \& Raymond, S. N. 2009, \apj, 696, 1854
\bibitem [{{Genzel} \& {Stutzki}(1989)}] {Genzel1989} Genzel, R., \& Stutzki, J. 1989, Ann. Rev. Astron. Astrophys., 27, 41
\bibitem [{{Ghosh} \& {McSween}(1998)}] {Ghosh1998} Ghosh, A. \& McSween, H. Y. 1998, Icar., 134, 187
\bibitem [{{Gilli} {et~al.}(2006)}] {Gilli2006} Gilli, T., Israelian, G., Ecuvillon, A., Santos, N. C., \& Mayor, M. 2006, \aap, 449, 723
\bibitem [{{Gounelle} \& {Meynet}(2012)}] {Gounelle2012} Gounelle, M., \& Meynet, G. 2012, \aap, 545, A4
\bibitem [{{Gritschneder} {et~al.}(2012)}] {Gritschneder2012} Gritschneder, M., Lin, D.N.C., Murray, S., D., Yin, Q.-Z., \& Gong, M.-N. 2012, \apj, 745, 22
\bibitem [{{Hanghoj} {et~al.}(2010)}] {Hanghoj2010} Hanghoj, K. Keleman, P. B., Hassler, D., \& Godard, M. 2010, J. Petrology, 51, 201
\bibitem [{{Hartmann} (2009)}] {Hartmann2009} Hartmann, L. 2009, Accretion Processes in Star Formation (2nd ed.; Cambridge: Cambridge Univ. Press)
\bibitem [{{Jacobsen} {et~al.}(2008)}] {Jacobsen2008} Jacobsen, B., Yin, Q.-z., Moynier, F., et al., E\&PSL, 272, 353
\bibitem [{{Jura}(2003)}] {Jura2003} Jura, M. 2003, \apj, 584, L91
\bibitem [{{Jura}(2006)}] {Jura2006} Jura, M. 2006, \apj, 653, 613
\bibitem [{{Jura} \& {Xu}(2012)}] {Jura2012a} Jura, M., \& Xu, S. 2012, \aj, 143, 6
\bibitem [{{Jura} \& {Xu}(2013)}] {Jura2013} Jura, M., \& Xu, S. 2013, \aj, 145, 30
\bibitem [{{Jura} {et~al.} (2012b)}] {Jura2012b} Jura, M., Xu, S., Klein, B., Koester, D., \& Zuckerman, B. 2012, \apj, 750, 69
\bibitem [{{Jura} \& {Young}(2014)}] {Jura2014} Jura, M. \& Young, E. D. 2014, Ann. Rev. Earth Planet. Sci., 42, in press
\bibitem [{{Klein} {et~al.} (2010)}] {Klein2010} Klein, B., Jura, M., Koester, D., Zuckerman, B., \& Melis, C. 2010, \apj, 709, 650
\bibitem [{{Larsen} {et~al.}(2011)}] {Larsen2011} Larsen, K., Trinquier, A., Paton, C., et al. 2011, \apj, 735, L37
\bibitem [{{Liu} {et~al.}(2012)}] {Liu2012} Liu, M.-C., Chaussidon, M., Gopel, G., \& Lee, T. 2012, E\&PSL, 327, 75
\bibitem [{{Lodders} (2003)}] {Lodders2003} Lodders, K. 2003, \apj, 591, 1220
\bibitem [{{Martin} {et~al.}(2009)}] {Martin2009} Martin, P. J., Knoedlseder, J., Diehl, R., \& Meynet, G. 2009, \aap, 506, 703
\bibitem [{{McSween} \& {Huss}(2010)}] {McSween2010} McSween, H. Y., \& Huss, G. R. 2010, Cosmochemistry (Cambridge: Cambridge Univ. Press)
\bibitem [{{Meyer} \& {Clayton}(2000)}] {Meyer2000} Meyer, B. S., \& Clayton, D. D. 2000, Space Sci. Rev., 92, 133
\bibitem [{{Pan} {et~al.}(2012)}] {Pan2012} Pan, L., Desch, S. J., Scannapieco, E., \& Timmes, F. X. 2012, \apj, 756, 102
\bibitem [{{Prantzos} (2004)}] {Prantzos2004} Prantzos, N. 2004, \aap, 420, 1033
\bibitem [{{Presnall} \& {Hoover}(1987)}] {Presnall1987} Presnall, D. C., \& Hoover, J. D. in Magmatic Processes: Physicochemical Principles, the Geochemical Society Special Publications, 1, 75
\bibitem [{{Scott} {et~al.}(2001)}] {Scott2001} Scott, E. R. D., Haack, H., \& Love, S. G. 2001, M\&PS, 36, 869
\bibitem [{{Tang} \& {Dauphas}(2012)}] {Tang2012} Tang, H., \& Dauphas, N. 2012, E\&PSL, 359,248
\bibitem [{{Turcotte} \& {Schubert}(2002)}] {Turcotte2002} Turcotte, R., \& Schubert, G. 2002, Geodynamics (2nd ed.; Cambridge: Cambridge Univ. Press)
\bibitem [{{Vasileiadis} {et~al.}(2013)}] {Vasileiadis2013} Vasileiadis, A., Nordlund, A., \& Bizarro, M. 2013, \apj, 769, L8
\bibitem [{{Voss} {et~al.}(2010)}] {Voss2010} Voss, R., Diehl, R., Vink, J. S., \& Hartmann, D. H. 2010, \aap, 520, A51
\bibitem [{{Wang} {et~al.}(2007)}] {Wang2007} Wang, W., Harris, J., Diehl, R., et al., 2007, \aap, 469, 1005
\bibitem [{{Warren}(2005)}] {Warren2005} Warren, P. H. 2005, Meteorit. Planet. Sci., 40, 477
\bibitem [{{Wasson} \& {Kallemeyn} (1988)}] {Wasson1988} Wasson, J. T., \& Kallemeyn, G. W. 1988, Phil. Trans. R. Roc. A, 325, 535
\bibitem [{{Xu} \& {Jura} (2012)}] {Xu2012} Xu, S. \& Jura, M. 2012, \apj, 745, 88
\bibitem [{{Xu} {et~al.}(2013)}] {Xu2013} Xu, S., Jura, M., Koester, D., Klein, B., \& Zuckerman, B. 2013, \apj, 766, 132
\bibitem [{{Zhou} {et~al.}(2013)}] {Zhou2013} Zhou, Q., Yin, Q.-Z., Young, E. D., et al. 2013, GeCoA, 110, 152
\bibitem [{{Zuckerman} {et~al.}(2011)}] {Zuckerman2011} Zuckerman, B., Koester, D., Dufour, P., et al. 2011, \apj, 739, 101
\bibitem [{{Zuckerman} {et~al.}(2007)}] {Zuckerman2007} Zuckerman, B., Koester, D., Melis, C., Hansen, B., \& Jura, M. 2007, \apj, 671, 872
\bibitem [{{Zuckerman} {et~al.}(2010)}] {Zuckerman2010} Zuckerman, B., Melis, C., Klein, B., Koester, D., \& Jura, M. 2010, \apj, 722, 725
 \end{thebibliography}

\end{CJK}
\end{document}